\documentclass[preprint,12pt]{elsarticle}

\usepackage{amssymb}

\journal{Nucl.\ Instrum.\ Methods Phys.\ Res.\ A}

\begin{document}

\begin{frontmatter}



\title{Impact of motion along the field direction on geometric-phase-induced
false electric dipole moment signals}


\author{H.\ Yan}
\author{B.\ Plaster}
\address{Department of Physics and Astronomy,
University of Kentucky, Lexington, KY, 40506 USA}

\begin{abstract}
Geometric-phase-induced false electric dipole moment (EDM) signals,
resulting from interference between magnetic field gradients and
particle motion in electric fields, have been studied extensively in
the literature, especially for neutron EDM experiments utilizing
stored ultracold neutrons and co-magnetometer atoms.  Previous studies
have considered particle motion in the transverse plane perpendicular
to the direction of the applied electric and magnetic fields.  We
show, via Monte Carlo studies, that motion along the field direction
can impact the magnitude of this false EDM signal if the wall surfaces
are rough such that the wall collisions can be modeled as diffuse,
with the results dependent on the size of the storage cell's dimension
along the field direction.
\end{abstract}


\end{frontmatter}



\section{Introduction}

If a non-zero neutron electric dipole moment (EDM) is observed, such a
discovery would provide evidence for parity-violation and
time-reversal-symmetry-violation beyond that of the standard model
\cite{pospelov05}.  Neutron EDM searches are usually based on a
Nuclear Magnetic Resonance (NMR) technique, in which the Larmor
precession frequencies are compared for parallel and anti-parallel
(weak) magnetic and (strong) electric field configurations.  A value
for, or a limit on, the EDM is then extracted from the frequency
difference for these two field configurations.  All recent experiments
have been performed with ultracold neutrons (UCN), neutrons with
speeds less than $\lesssim 7$ m/s, or energies $\lesssim 350$ neV
\cite{ucn_book}.  Their low speeds permit storage in cells for long
periods of time (in principle, up to the $\beta$-decay lifetime), and
reduce systematic errors related to the neutron velocity.  Of crucial
importance to all EDM searches is monitoring of the magnetic field.
This is typically accomplished using in-situ co-magnetometer and/or
external magnetometer atoms, such as $^{199}$Hg \cite{baker06},
$^{3}$He \cite{golub94,lamoreaux09}, and $^{133}$Cs \cite{altarev09}.

A systematic error that has been discussed extensively in the
literature recently
\cite{pendlebury04,lamoreaux05,barabanov06,golub08} is the so-called
geometric phase effect, resulting from interference between magnetic
field gradients and the $(\vec{E} \times \vec{v})/c^2$ motional
magnetic field in the particle rest frame.  The result is a frequency
shift $\delta\omega$ proportional to the electric field $E$, for both
the UCN and magnetometer atoms, which is dependent on the particles'
(geometric) trajectories.  Because the frequency shift is
proportional to $E$, the effect could be interpreted as a false EDM
signal.  A general description of this effect based on the density
matrix formalism was developed \cite{lamoreaux05,barabanov06,golub08},
in which the frequency shift was shown to be proportional to the
Fourier transformation of the velocity autocorrelation function for a
constant gradient $\partial B_{0z} / \partial z$ (assuming fields
along the $z$-axis).  The frequency shift $\delta\omega$ was shown to
be of the form \cite{lamoreaux05,barabanov06,golub08}
\begin{equation}
\delta\omega = \frac{\gamma^{2}}{4}\frac{\partial B_{0z}}{\partial
z}\frac{E}{c^{2}}\int^{t}_{0} d\tau R(\tau) \cos\omega_0\tau
\end{equation}
where the autocorrelation function is
\begin{eqnarray}
R(\tau) &=& \langle y(t)v_{y}(t-\tau) + x(t)v_{x}(t-\tau) \nonumber \\
&& -~y(t-\tau)v_{y}(t) - x(t-\tau)v_{x}(t) \rangle \nonumber \\
&=& 2 {\int^\tau_0} \langle v_y(t) v_y(t-\tau) +
v_x(t) v_x(t-\tau) \rangle dt.
\label{eq:correlation-function}
\end{eqnarray}
Here, $\gamma$ is the gyromagnetic ratio, and $\omega_0 = \gamma
B_{0z}$ is the nominal precession frequency for a magnetic field
$B_{0z}$.

\section{Motion Along the Field Direction}
As was pointed out in \cite{lamoreaux05}, both wall and gas collisions
can lead to a suppression of the autocorrelation function, and thus of
the frequency shift.  However, upon inspection of Eq.\
(\ref{eq:correlation-function}), it can be seen that if we consider
three-dimensional motion, the velocity component along the $z$-axis,
$v_z$, may impact the autocorrelation function, even though $v_z$
itself does not contribute to the $(\vec{E} \times \vec{v})/c^2$
field.  The reasoning is as follows.  Suppose a particle undergoes a
collision with a ``$z$-wall''.  If the wall collision is specular, the
velocity components before ($v$) and after ($v'$) the collision will
transform according to
\begin{equation}
v'_x = v_x,~~~~~v'_y = v_y,~~~~~v'_z = -v_z,
\end{equation}
and thus specular wall collisions will have no impact on the
autocorrelation function $R(\tau)$.  However, for collisions with
rough wall surfaces, such that the reflection angles can be modeled
as Lambertian diffuse,
the $v_x$ and $v_y$ velocity components before and after
the collision will not, in general, be correlated
\begin{equation}
v'_x \neq v_x,~~~~~v'_y \neq v_y,~~~~~
v'_z \neq -v_z.
\end{equation}

This suggests that an understanding of the geometric-phase false EDM
effect in future experiments will require detailed studies of the
autocorrelation functions of the stored UCN and co-magnetometer atoms,
and that these functions will depend on the type of wall collisions
the stored particles undergo, and on the size of the storage volume
along the field direction (thus affecting the wall collision rate).
We have employed the autocorrelation function approach of
\cite{lamoreaux05,barabanov06,golub08} in Monte Carlo studies of the
impact of particle motion along the field direction (i.e., velocities
along the $z$-axis) on the magnitude of the geometric-phase-induced
false EDM signal.  In the next section, we show several illustrative
results from our Monte Carlo studies of this effect.

\begin{figure}[t]
\begin{center}
\includegraphics[scale=0.45]{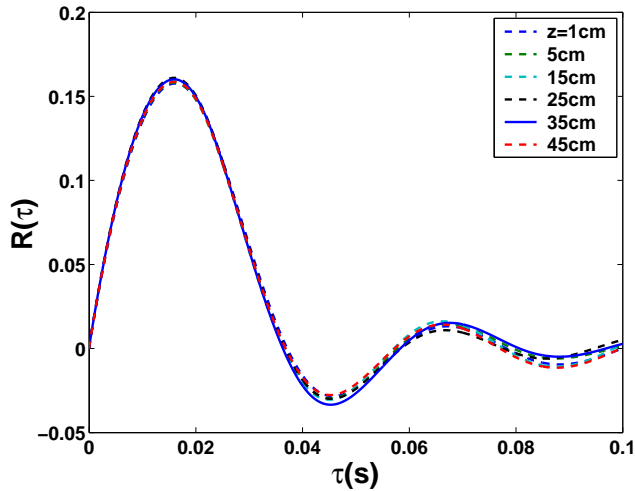}
\caption{(Color online) Results from simulations of the
autocorrelation function $R(\tau)$ for UCN undergoing specular wall
collisions in a 10~cm $\times$ 10 cm two-dimensional square geometry,
for the different indicated $z$-dimensions.}
\label{fig.1}
\end{center}
\end{figure}

\begin{figure}[t]
\begin{center}
\includegraphics[scale=0.34,clip=]{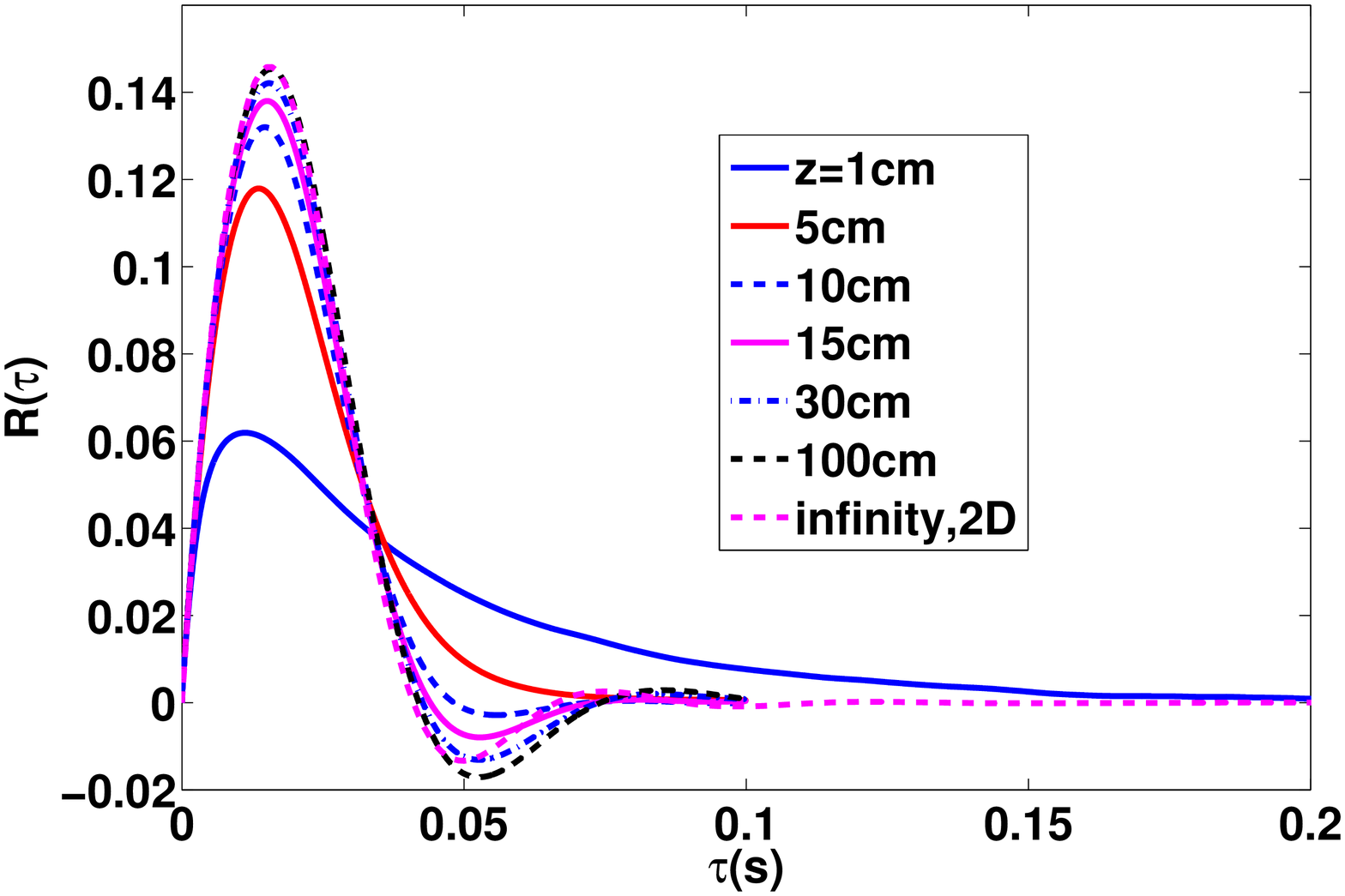}
\caption{(Color online) Results from simulations of the
autocorrelation function $R(\tau)$ for UCN undergoing Lambertian
diffuse wall collisions in a 10~cm $\times$ 10 cm two-dimensional
square geometry, for the different indicated $z$-dimensions.}
\label{fig.2}
\end{center}
\end{figure}

\begin{figure}[h]
\begin{center}
\includegraphics[scale=0.63, angle=0]{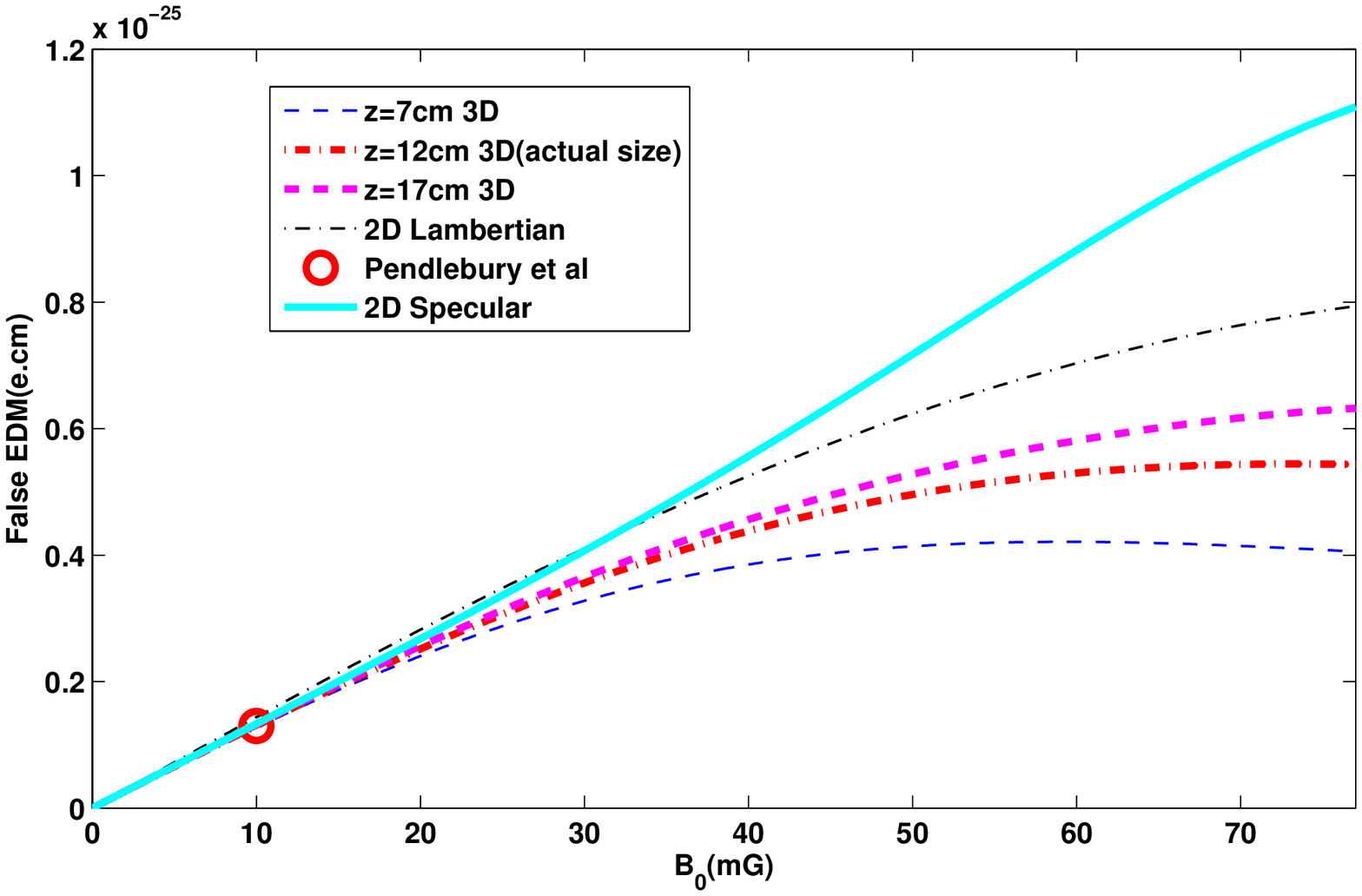}
\caption{(Color online) Results from simulations of the false EDM of
$^{199}$Hg co-magnetometer atoms stored in a cylindrical geometry with
a radius of 0.25~m, and for various three-dimensional heights.  The
fractional field gradient is $10^{-5}$ cm$^{-1}$.  The red circle
indicates the value for $d_{af\mathrm{Hg}}$ reported in
\cite{pendlebury04}.}
\label{fig.3}
\end{center}
\end{figure}

\section{Results from Monte Carlo Studies}
Monte Carlo codes were developed to study the impact of motion along
the field direction on the autocorrelation function.  Collisions with
the wall surfaces were modeled as either specular or Lambertian
diffuse (i.e., reflection
angles sampled from a $f(\theta)d\Omega \propto \cos\theta d\Omega$
angular distribution).  For UCN, we sampled a $v^2 dv$ velocity
distribution up to some cut-off velocity, and for co-magnetometer
atoms, we sampled a Maxwell-Boltzmann distribution.  Below we discuss
several example results from our Monte Carlo studies.

First, we demonstrate in Fig.\ \ref{fig.1} that the autocorrelation
function is, indeed, insensitive to the three-dimensional size of the
storage volume for purely specular wall collisions.  This figure shows
plots of $R(\tau)$ for UCN stored in a 10 cm $\times$ 10 cm
two-dimensional square geometry, for different $z$-dimensions.
Indeed, there is little difference between these results.
Thus, for specular wall collisions, simulations in two-dimensions
are sufficient.

Second, Fig.\ \ref{fig.2} shows results from simulations for wall
collisions modeled as purely Lambertian diffuse.  Here, it can be seen
that as the $z$-dimension increases, the results approach the limiting
two-dimensional result in which the collision rate with the $z$-walls
becomes small relative to the collision rate with the walls in the
transverse $x$-$y$ plane.  In the opposite limit, as the $z$-dimension
becomes small relative to the transverse $x$-$y$ dimensions, the
collision rate with the $z$-walls increases, and the autocorrelation
function becomes highly suppressed.

Third, as a final example, we considered the false EDM of the
$^{199}$Hg co-magnetometer atoms stored in the cylindrical geometry of
the neutron EDM experiment reporting the current benchmark upper limit
of $<2.9 \times 10^{-26}$ $e$-cm (90\% C.L.)  \cite{baker06}.  As
reported in \cite{pendlebury04}, for a gradient of $\partial B_{0z} /
\partial z = 1$ nT/m at a field of $B_{0z} = 1$ $\mu$T, or a
fractional gradient of $(\partial B_{0z} / \partial z)/B_{0z} =
10^{-5}$ cm$^{-1}$, the false EDM signal of the Hg atoms was (using
their notation) $d_{af\mathrm{Hg}} = 1.3 \times 10^{-26}$ $e$-cm.

Our simulations assumed the $^{199}$Hg atoms were at a temperature of
300~K (with no buffer gas, thus undergoing wall collisions only), the
radius of the cylindrical geometry was 0.25~m (as reported in
\cite{pendlebury04}), and the nominal three-dimensional size (height)
of the cylidrical geometry was 0.12 m (as reported in
\cite{harris06}).  Some results of our simulations are shown in Fig.\
\ref{fig.3}, where we have plotted the false EDM signal of the
$^{199}$Hg atoms, $d_{af\mathrm{Hg}}$, versus the magnetic field
$B_{0z}$.  As a consistency check, we note the good agreement at
$B_{0z} = 1$ $\mu$T between the result obtained in our simulations,
and the value of $1.3 \times 10^{-26}$ $e$-cm reported in
\cite{pendlebury04}.  The electric field was assumed to be $E = 10$
kV/cm \cite{baker06}.

At the operating magnetic field of the previous experiment [$B_{0z} =
1$ $\mu$T (= 10 mG)], the differences between the two-dimensional case
and the three-dimensional case are much smaller than the extracted EDM
limit.  However, as the value of the magnetic field increases (i.e.,
as the precession frequency becomes large relative to any angular
velocity of the stored particles), the differences between the
two-dimensional case and the three-dimensional case become apparent,
with the magnitude of the false EDM showing a significant dependence
on the exact size of the dimension along the $z$-axis (as demonstrated
for the two other, arbitrarily chosen, different $z$-dimension sizes).

\section{Discussion}

In summary, our Monte Carlo studies suggest that in neutron EDM
experiments with stored UCN and co-magnetometer atoms, if the wall
surfaces are rough such that the collisions are modeled as Lambertian
diffuse, the magnitude of the geometric-phase-induced false EDM signal
will be dependent on the size of the storage cell's dimension along
the field direction.  Indeed, as expected intuitively, the magnitude
of the false EDM can be suppressed by utilizing a geometry with a
relatively small dimension along the field direction.  For planning of
future neutron EDM experiments, our studies suggest that it will be
important to understand all possible physical processes which may
impact the autocorrelation function, including quantifying the degree
to which the wall collisions can be modeled as either specular or
diffuse, as presented in this work.  Indeed, a complete understanding
of the geometric-phase-induced false EDM signal in future experiments
will require experimental measurements of the autocorrelation
function, coupled to three-dimensional simulations. \\

\noindent\textbf{Acknowledgments} \\ \\
We thank B.\ Filippone and R.\ Golub for valuable discussions.  This
work was supported in part by the U.S.\ Department of Energy under
award number DE-FG02-08ER41557, and by the University of Kentucky.








\end{document}